\documentclass[showpacs,amsmath,amssymb,twocolumn,superscriptaddress,
aps,prb]{revtex4}
\usepackage{graphicx}

\begin{document}
\title{Magnetic-field induced quantum-size cascades in superconducting
nanowires}
\author{A. A. Shanenko}
\affiliation{TGM, Departement Fysica, Universiteit Antwerpen, B-2020
Antwerpen, Belgium}
\affiliation{Bogoliubov Laboratory of Theoretical Physics,
Joint Institute for Nuclear Research, 141980 Dubna, Russia}
\author{M. D. Croitoru}
\affiliation{EMAT, Departement Fysica, Universiteit Antwerpen, B-2020
Antwerpen, Belgium}
\author{F. M. Peeters}
\affiliation{TGM, Departement Fysica, Universiteit Antwerpen, B-2020
Antwerpen, Belgium}

\date{\today}
\begin{abstract}
In high-quality nanowires, quantum confinement of the transverse
electron motion splits the band of single-electron states in a
series of subbands. This changes in a qualitative way the scenario
of the magnetic-field induced superconductor-to-normal transition.
We numerically solve the Bogoliubov-de Gennes equations for a clean
metallic cylindrical nanowire at zero temperature in a parallel
magnetic field and find that for diameters $D\lesssim 10\div15\,
{\rm nm}$, this transition occurs as a {\it cascade} of subsequent
jumps in the order parameter~(this is opposed to the smooth second-%
order phase transition in the mesoscopic regime). Each jump is
associated with the depairing of electrons in one of the single-%
electron subbands. As a set of subbands contribute to the order
parameter, the depairing process occurs as a cascade of jumps.
We find pronounced quantum-size oscillations of the critical
magnetic field with giant resonant enhancements. In addition
to these orbital effects, the paramagnetic breakdown of Cooper
pairing also contributes but only for smaller diameters, i. e.,
$D\lesssim 5\,{\rm nm}$.
\end{abstract}

\maketitle

\section{Introduction}
\label{sec1}

High-quality superconducting nanostructures as, e.g., single-crystal ${\rm
Sn}$ nanowires~\cite{tian}, polycrystalline (but made of strongly coupled
grains) ${\rm Al}$ nanowires~\cite{arut,altom} and single-crystalline
atomically uniform ${\rm Pb}$ nanofilms~\cite{guo,eom,ozer,ozer1} have
recently been fabricated. It was possible to minimize the disorder such
that the electron mean free path was about or larger than the specimen
thickness.~\cite{arut,altom,ozer1} In this case the scattering on
nonmagnetic imperfections influences only the electron motion parallel
to the wire/film, while the perpendicular electron motion is governed by
the transverse-size quantization. Indeed, photoemission spectra of
ultrathin single-crystal ${\rm Pb}$ films showed clear signatures of
the splitting  of the electron band into a series of subbands due to
the transverse-size quantization.~\cite{guo} In the presence of minimal
disorder the so-called Anderson theorem~\cite{degen} (see, also,
discussion in Ref.~\onlinecite{skvorts}) controls the effect of
nonmagnetic impurities. Thus, one can expect that the study of a clean
system with quantized transverse electron motion can capture important
issues concerning the impact of quantum confinement on the
superconducting characteristics in high-quality nanowires/nanofilms.

The single-electron subbands appearing due to the transverse quantization,
move in energy with changing specimen thickness. When the bottom of a
subband passes through the Fermi surface, the density of single-electron
states at the Fermi level increases abruptly. This results in size-dependent
superconducting resonances~\cite{blatt} and in quantum-size oscillations of
the superconducting properties as function of the thickness. Recently such
quantum-size oscillations in the critical temperature $T_c$ of superconducting
${\rm Pb}$ nanofilms were observed at a high level of experimental precision
and sophistication.~\cite{guo,eom} Quantum-size superconducting resonances
were shown to be responsible for an increase of the superconducting transition
temperature in ${\rm Al}$ and ${\rm Sn}$ nanowires with decreasing thickness~%
\cite{our1}.

The transverse quantization of the electron motion should strongly influence
the superconducting-to-normal phase transition driven by a magnetic field in
such high-quality nanowires/nanofilms. In the present paper we limit ourselves
to nanowires in a parallel magnetic field and ignore the vortex formation
because vortices cannot nucleate in very thin superconducting wires.

According to the Ginzburg-Landau (GL) theory~\cite{silin,lutes}, the critical
magnetic field is expected to increase as $1/D$ in the Meissner state, with
$D$ the diameter of the mesoscopic wire. Furthermore, the superconducting-%
to-normal phase transition in a magnetic field is of second order for such
mesoscopic wires while being of first order in bulk~(for type I
superconductors)~\cite{degen}. It is a general characteristic of the GL
theory that this transition becomes of second order in mesoscopic specimens~%
\cite{degen,peeters}. Recent calculations based on the Bogoliubov-de Gennes
(BdG) equations~\cite{han} for wires with diameters $20\div200\,{\rm nm}$
confirmed the GL result and revealed a smooth superconducting-to-normal
transition in  a parallel magnetic field at any temperature below $T_c$.
This is in agreement with recent experimental data on ${\rm Sn}$~%
\cite{tian,jank} and ${\rm Zn}$~\cite{tian2} nanorods. Hence, one may
conclude that effects of the transverse quantization of the electron
motion are not significant for metallic superconducting wires with width
larger than $20\,{\rm nm}$.

In the present paper we show that the situation changes dramatically for
smaller widths. Our analysis is based on a numerical self-consistent
solution of the BdG equations for a clean cylindrical metallic nanowire.
We predict that at zero temperature the superconducting-to-normal
transition driven by a magnetic field parallel to the nanowire, occurs
as a {\it cascade} of jumps in the order parameter (with clear signatures
of hysteretic behavior) for diameters $D\lesssim 10\div15\,{\rm nm}$. This
qualitative change is accompanied by pronounced quantum-size oscillations
of the critical magnetic field with large enhancements at the points of the
superconducting resonances. In addition to these orbital effects, we found
that Pauli paramagnetism can also contribute but its role is only significant
for smaller diameters, i.e., $D\lesssim 5\,{\rm nm}$.

\section{Bogoliubov-de Gennes equations}
\label{sec2}

In the clean limit the BdG equations~\cite{degen} read
\begin{subequations}
\begin{equation}
E_n u_n({\bf r})={\widehat H}_e\,u_n({\bf r}) +
                                   \Delta({\bf r})\,v_n({\bf r}),
\label{BdG1}
\end{equation}
\begin{equation}
E_n v_n({\bf r})=\Delta^*({\bf r})\,u_n({\bf r})-
                                  {\widehat H}^*_e\,v_n({\bf r}),
\label{BdG2}
\end{equation}
\end{subequations}
where $\Delta({\bf r})$ stands for the superconducting order parameter
($*$ for complex conjugate), $E_n$ is the quasiparticle energy, $u_n({\bf
r})$ and $v_n({\bf r})$ are the particle-like and hole-like wave functions.
The single-electron Hamiltonian appearing in Eqs.~(\ref{BdG1}) and
(\ref{BdG2}) is given by
\begin{equation}
{\widehat H}_e=\frac{1}{2m_e}\left(-i\hbar\nabla-
                                       \frac{e}{c}{\bf A}\right)^2-E_F,
\label{He}
\end{equation}
with $m_e$ the electron band mass~(can be set to the free-electron mass
without loss of generality), and $E_F$ the Fermi level. The BdG should
be solved in a self-consistent manner, together with the self-%
consistency relation
\begin{equation}
\Delta({\bf r}) = g\sum_n\,u_n({\bf r}) v_n^*({\bf r})\bigl(1-2f_n\bigr),
\label{self}
\end{equation}
with $g$ the coupling constant and $f_n=f(E_n)$ the Fermi function~%
\cite{degen}.

An important issue is the range of the states included in the sum in
Eq.~(\ref{self}). The usual prescription concerns the quasiparticles
with positive energies $E_n$. At the same time the corresponding
single-electron energy $\xi_n$ should be located in the Debye window,
$|\xi_n| < \hbar\omega_D$ with $\omega_D$ the Debye frequency and
\begin{equation}
\xi_n = \int\!{\rm d}^3r \left[u^*_n({\bf r}){\widehat H}_e
      u_n({\bf r})+v^*_n({\bf r}){\widehat H}_e v_n({\bf r})\right].
\label{single}
\end{equation}
However, in the presence of a magnetic field, this prescription is
modified: ${\widehat H}_e|_{{\bf A}=0}$ is used rather than ${\widehat
H}_e$ in Eq.~(\ref{single}). It is well-known that the selection
$|\xi_n| < \hbar\omega_D$ appears as a result of the delta-function
approximation for the effective electron-electron interaction. Such
an approximation neglects a complex structure of the Fourier transform
of the pair interaction. The problem is cured by the well-known cut-%
off in {\it the canonical-momentum} space. Such a cut-off results in
the above selection rule for $\xi_n$ with ${\widehat H}_e$ replaced
by ${\widehat H}_e |_{{\bf A}=0}$~(see, for instance, Refs.~%
\onlinecite{degen} and \onlinecite{han}). Second, the requirement of
positive quasiparticle energies has to be weakened in the presence of
a magnetic field. Namely, one needs to include the states having
positive quasiparticle energies only at zero magnetic field. This
allows one to investigate also the regime of gapless superconductivity
when the presence of quasiparticles with negative energies manifests
the depairing reconstruction of the ground state~(see Eq.~(\ref{ansatz})
below and Appendix~\ref{app}).

Due to transverse quantum confinement we set
\begin{equation}
u_n({\bf r})|_{{\bf r}\in S} = v_n({\bf r})|_{{\bf r}\in S}=0
\label{bound}
\end{equation}
on the wire surface. Periodic boundary conditions are used along the
nanowire. Screening of the external magnetic field can be neglected
for narrow wires. Then, for a constant magnetic field parallel to
the nanocylinder, $H_{||}$, it is convenient to use the well-known
Coulomb gauge. Thus, for cylindrical wires we have $\Delta({\bf r})
=\Delta(\rho)$ with $\rho,\varphi,z$ the cylindrical coordinates
(below the order parameter is chosen as a real quantity). The set
of relevant quantum numbers is $n=\{j,m,k\}$, with $j$ the quantum
number associated with $\rho$, $m$ the azimuthal quantum number,
and $k$ the wave vector of the quasi-free electron motion along
the nanowire. In this case the particle-like and hole-like wave
functions can be represented as
\begin{subequations}
\begin{equation}
u_n({\bf r})=u_{jmk}(\rho)\,
\frac{e^{\imath m \varphi}}{\sqrt{2\pi}}\,\frac{e^{ikz}}{\sqrt{L}},
\label{u}
\end{equation}
\begin{equation}
v_n({\bf r})=v_{jmk}(\rho)\,\frac{e^{\imath m
\varphi}}{\sqrt{2\pi}}\, \frac{e^{ikz}}{\sqrt{L}}, \label{v}
\end{equation}
\end{subequations}
with $L$ the length of the nanowire. Inserting Eqs.~(\ref{u}) and
(\ref{v}) into the BdG equations (\ref{BdG1}) and (\ref{BdG2}) and
using an expansion in terms of the Bessel functions~(see details in
Ref.~\onlinecite{sc}), the problem is reduced to the diagonalization
of a matrix.

\section{Discussion of numerical results}
\label{sec3}
\subsection{Resonances in the critical magnetic field}

\begin{figure*}
\resizebox{1.5\columnwidth}{!}
{\rotatebox{0}{\includegraphics{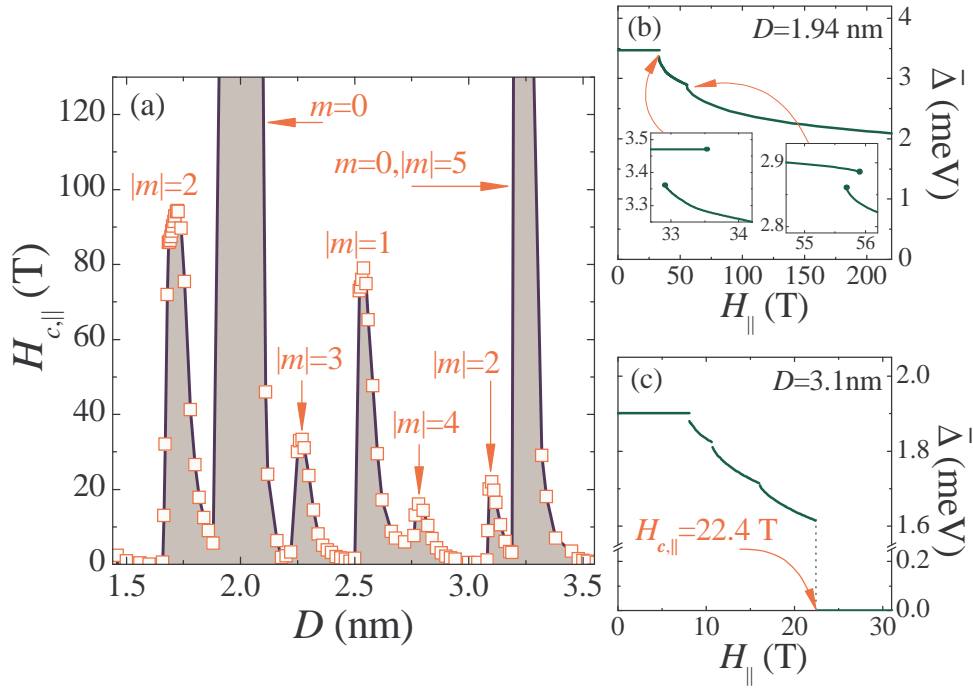}}}
\caption{(Color online) (a) Critical parallel magnetic field $H_{c,||}$
versus the nanowire diameter $D$, and spatially averaged order parameter
$\bar{ \Delta}$ as function of $H_{||}$ for the resonant diameters (b)
$D= 1.94\,{\rm nm}$~[governed by $(j,|m|)=(1,0)$] and (c) $D=3.1\,{\rm
nm}$~[governed by $(j,|m|)=(1,1)$].}
\label{fig1}
\end{figure*}

At a superconducting resonance the main contribution to the different
superconducting quantities comes from the subband (or subbands) whose
bottom passes through the Fermi surface. For cylindrical wires, the
subbands with the same $|m|$ are degenerate for $H_{||}=0$ and, hence,
any size-dependent resonant enhancement of the order parameter (e.g.,
the energy gap and the critical temperature) can be specified by the
set $(j,|m|)$ in the absence of a magnetic field. Due to quantum-size
oscillations in the pair-condensation energy, we get corresponding
oscillations in the critical magnetic field whose resonant
enhancements can also be labeled by $(j,|m|)$. Figure~\ref{fig1}(a)
shows the critical field $H_{c,||}$ calculated self-consistently
from Eqs.~(\ref{BdG1}) and (\ref{BdG2}) at zero temperature ($T=0$)
for an aluminum nanocylinder with diameter $D$. Note that $H_{c,||}$
is set as the magnetic field above which the spatially averaged order
parameter $\bar{\Delta}$ drops below $0.01\Delta_{\rm bulk}$, with
$\Delta_{\rm bulk}$ the bulk gap. Here, we consider as an example
${\rm Al}$ and take~\cite{degen} $\hbar\omega_D = 32.31\,{\rm meV}$
and $gN(0)=0.18$, where $N(0)$ stands for the bulk density of states.
For this choice $\Delta_{\rm bulk}=0.25\,{\rm meV}$. The effective
Fermi level is set to $E_F=0.9\,{\rm eV}$, which is used together with
the BdG equations within the parabolic band approximation~\cite{our2}.
As seen, $H_{c,||}$ exhibits huge enhancements as compared to the bulk
critical magnetic field $H_{c,{\rm bulk}}=0.01\,{\rm T}$~(to simplify
our discussion, we show first the results for extremely narrow quantum
wires). Resonances in $H_{c,||}$ are found to be very dependent on $D$
and $|m|$. The states with large $|m|$ are more strongly influenced
by $H_{||}$ and, so, the resonances in $H_{c,||}$ governed by large
$|m|$ are, as a rule, less pronounced. In contrast, the resonances
controlled by $m=0$ are very stable. For instance, a superconducting
solution to Eqs.~(\ref{BdG1}) and (\ref{BdG2}) exists at $D=1.94\,
{\rm nm}$ [the resonance associated with $(j,|m|)=(1,0)]$ even for
an abnormally large magnetic field of about $1000\,{\rm T}$. Similar
behavior is found for the resonance at $D=3.21\,{\rm nm}$ with $(j,m)
=(2,0)$. Note that in Fig.~\ref{fig1} two neighboring resonances with
$(j,m)=(2,0)$~[$D =3.21\,{\rm nm}$] and $(j,|m|)=(1,5)$ [$D=3.28\,
{\rm nm}$] merge and result in one profound increase in $H_{c,||}$.

\subsection{Quantum-size cascades}

\begin{figure*}
\resizebox{2.0\columnwidth}{!}{\rotatebox{0}{\includegraphics%
{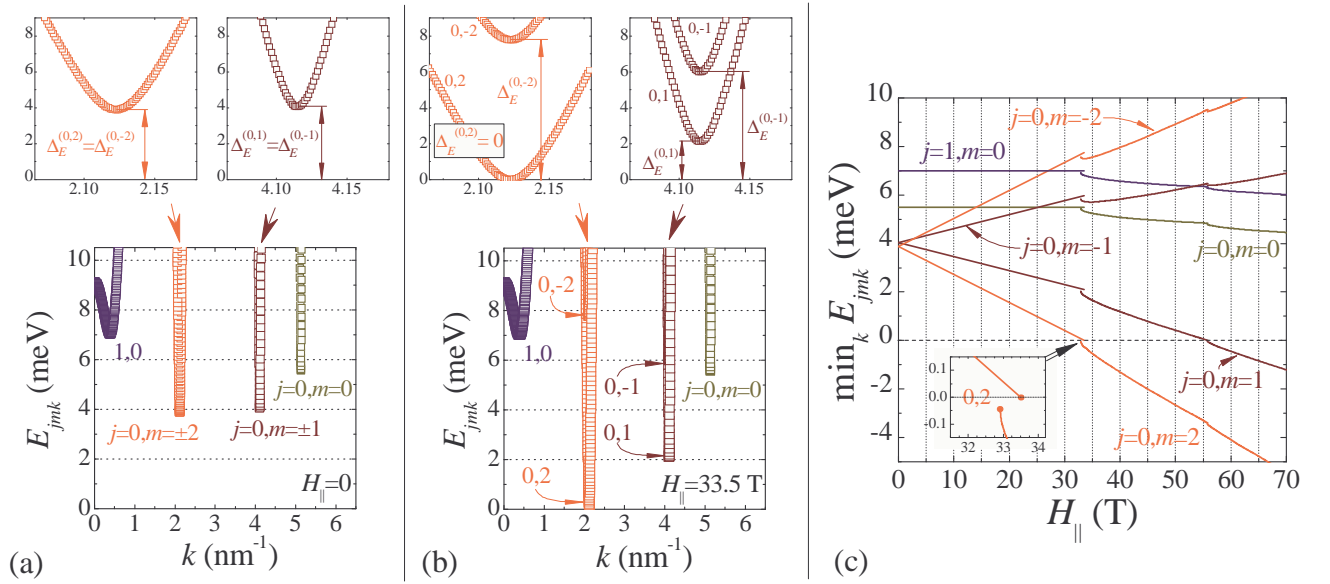}}}\caption{(Color online) The quasiparticle
energies $E_{jmk}$ versus $k$ for the four relevant branches
$(j,m)=(0,0),(0,\pm1),(0,\pm2)$ and $(1,0)$ at (a) $H_{||}=0$
and (b) $H_{||}=33.5\,{\rm T}$ for the resonant diameter $D=1.
94\,{\rm nm}$. In this case there are two nonzero solutions of
the BdG equations for $32.9\,{\rm T}\leq H_{||}\leq 33.5\,{\rm T}$
[see the left-side inset in Fig.~\ref{fig1}(b)], the upper one
disappears at $H_{||}=33.5\,{\rm T}$~[panel (b)] when the
quasiparticle branch $(0,2)$ touches zero. (c) The quantity
$\min_{\bf k}E_{jmk}$ versus $H_{||}$ for the different
quasiparticle branches at $D=1.94\,{\rm nm}$.}
\label{fig2}
\end{figure*}

Figures~\ref{fig1}(b) and \ref{fig1}(c) show two typical examples
($m=0$ and $|m|\not=0$) of how the spatially averaged order parameter
$\bar{\Delta}$ depends on $H_{||}$. To discuss these results, we
remark that the quasiparticle energies can be well approximated by
\begin{equation}
E_{jmk}=\sqrt{\xi^2_{jmk} + \Delta^2_{jmk}} - m\mu_B H_{||},
\label{ener}
\end{equation}
where $\xi_{jmk}$ is the single-electron energy given by Eq.~%
(\ref{single}) (at $H_{||} = 0$), $\mu_B$ stands for the Bohr
magneton and
\begin{equation}
\Delta_{jmk} =\int\limits_0^R\!{\rm d}\rho\;\rho\;\Delta(\rho)
                    \Bigl[|u_{jmk}(\rho)|^2+|v_{jmk}(\rho)|^2\Bigr],
\label{djmk}
\end{equation}
the averaged value of the order parameter as seen by $jmk-$quasiparticles
($R=D/2$). Equation (\ref{djmk}) can be derived within Anderson's
approximate solution of the BdG equations~\cite{anders}. This approximate
solution implies that the particle-like and hole-like wave functions are
chosen to be proportional to the eigenfunctions of ${\widehat H}_e$~(for
details, see Appendix~\ref{app}). Note that the dependence of $\Delta_{%
jmk}$ on $k$ is found to be negligible: $\Delta_{jmk}=\Delta_{jm}$~(see
Eq.~(\ref{djm})). As follows from Eq.~(\ref{ener}), quasiparticles with $m
 > 0$ are moved down in energy by $H_{||}$. Each time when a quasiparticle
branch specified by a positive $m$ touches zero, a jump in $\bar{\Delta}$
occurs. When a branch controlling a resonant enhancement approaches zero,
$\bar{\Delta}$ jumps down to zero and the superconducting solution
disappears (see Fig.~\ref{fig1}(c)). Other quasiparticle branches are less
important (due to a smaller density of states) and are responsible for small
(sometimes almost insignificant) jumps in $\bar{\Delta}$. In particular, at
$D=1.94\,{\rm nm}$ the first small jump in $\bar{\Delta}$~(see Fig.~\ref%
{fig1}(b)) is located at $H_{||}=33.5\,{\rm T}$. Here the branch with $j=0,
m=2$ touches zero (see Fig.~\ref{fig2}). The insets in Fig.~\ref{fig1}(b)
show details of jumps in $\bar{\Delta}$. As seen, there are clear signatures
of hysteretic behavior: in the vicinity of any jump the BdG equations has
two possible solutions.

To properly clarify details of the hysteretic behavior, we performed a
numerical analysis for sufficiently large values of the unit-cell length $L$,
controlling periodic boundary conditions in the longitudinal direction. In
particular, the limit $L \to \infty$ can be approached only when $L > 10\div
20\,{\rm \mu m}$~($L/D > 10^5$). For $m=0$ the last term in Eq.~(\ref{ener})
is ``switched off" and, so, $\bar{\Delta}$ exhibits only a sequence of small
jumps for the resonances governed by $m=0$~[see Fig.~\ref{fig1}(b)]. For
any quasiparticle branch an energy gap $\Delta^{(jm)}_E$~(see Fig.~\ref%
{fig2}) can be introduced, and the total excitation energy gap is defined
as $\Delta_E=\min \Delta^{(jm)}_E$. Stress that in general, $\Delta_{jm}
\not=\Delta^{(jm)}_E$, only at $H_{||}=0$ we have $\Delta_E = \min\Delta_{jm}$.
Thus, a jump in $\bar{\Delta}$ appears when one of $\Delta^{(jm)}_E$ becomes
zero. In particular, the left-side inset in Fig.~\ref{fig1}(b) shows that
there exist two solutions in the interval from $H_{||}=32.94\,{\rm T}$ to
$H_{||}=33.5\,{\rm T}$~(the first jump in the order parameter as a function
of $H_{||}$). For the upper solution we have $\Delta_E=\Delta_{0,2}\geq 0$
that decreases linearly with $H_{||}$ until touching zero at $H_{||}=33.5
\,{\rm T}$~[see the quasiparticle energies corresponding to the upper
solution and given in Figs.~\ref{fig2}(a) and \ref{fig2}(b) for $H_{||}
=0$ and $H_{||}=33.5\,{\rm T}$, respectively]. For the lower solution
$\Delta^{(0,2)}_E=0$ and, so, the gapless regime is realized with $\Delta_E
=\Delta^{(0,2)}_E=0$. For more detail, Fig.~\ref{fig2}(c) shows how
$\min_{{\bf k}}E_{jmk}$ varies with $H_{||}$ for $D=1.94\,{\rm nm}$. As
seen, each relevant quasiparticle branch exhibits signatures of two small
jumps [corresponding to the jumps in the order parameter given in Fig.~%
\ref{fig2}(b)]. After the first jump $\min_{\bf k} E_{0,2,k}<0$~(see the
inset) and, so, $\Delta^{(0,2)}_E=0$. After the second jump (for $H_{||}
> 55.85\,{\rm T}$) $\min_{\bf k} E_{0,1,k}$ becomes negative and, hence,
we get $\Delta_E=\Delta^{(0,1)}_E=\Delta^{(0,2)}_E=0$. In the near vicinity
of the second jump, for $55.7\,{\rm T} \leq H_{||} \leq 55.85\,{\rm T}$, we
again find two nonzero solutions for the BdG equations: $\Delta^{(0,1)}_E\not
=0$ for the upper solution (except of the edge point $H_{||}=55.85\,{\rm nm}$)
and $\Delta^{(0,1)}_E =0$ for the lower one. Above, we discussed only numerical
results for the resonances. The same conclusions hold for the off-resonant
points. However, the eventual jump to zero in $\bar{\Delta}$ at $H_{||}= H_{c,
||}$ is, of course, much less pronounced in this case.

\begin{figure*}
\resizebox{1.4\columnwidth}{!}{\rotatebox{0}{\includegraphics{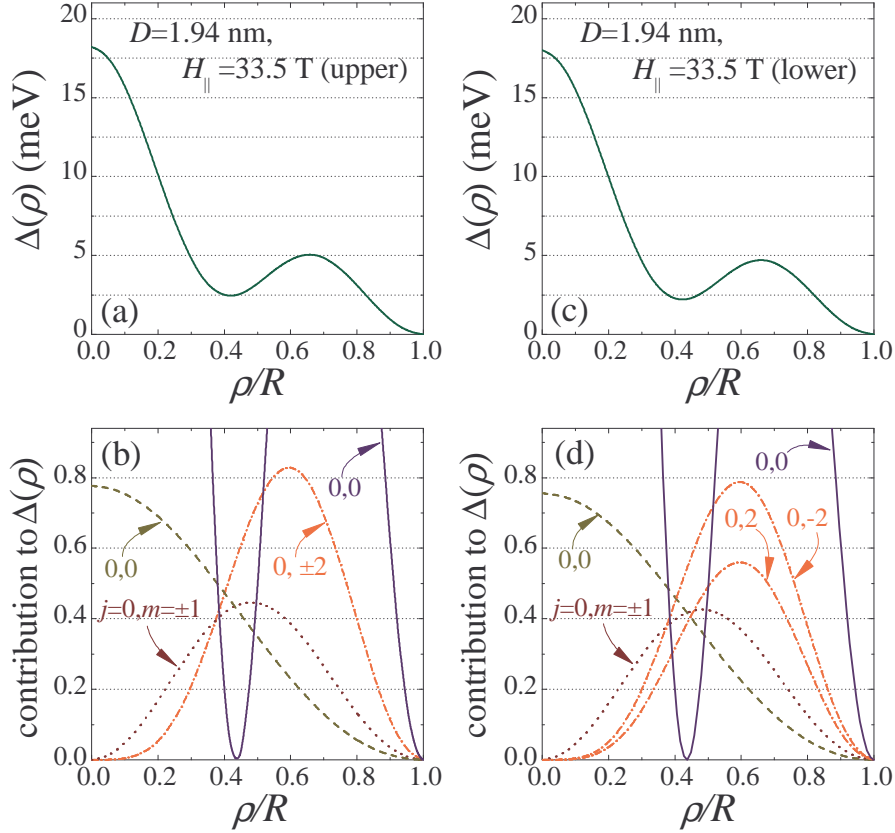}}}\caption{(Color online) The upper (a,b) and lower (c,d)
solutions of the BdG equations at $H_{||}=33.5\,{\rm T}$~(the resonant
diameter $D=1.94\,{\rm nm}$): (a) and (c), the order parameter $\Delta
(\rho)$; (b) and (d), the corresponding contribution of the different
relevant single-electron subbands.}
\label{fig3}
\end{figure*}

Now the question arises: {\it what is the physics underlying these
cascades of jumps in the order parameter?} A jump appears when one of
the relevant quasiparticle branches touches zero. From this point on,
such a branch "supplies" the system with states having negative
quasiparticle energies [see the discussion about Eq.~(\ref{self}) in
Sec.~\ref{sec2}]. For such quasiparticles $f_n=1$ at zero temperature
or, in other words, these quasiparticles survive even at $T=0$. It
means that we face a reconstruction of the ground state. To have a
feeling about such a reconstruction, let us consider the multiband
BCS ansatz for the ground-state wave function~(see Appendix~\ref{app},
Eq.~(\ref{mult})). This ansatz reads
\begin{equation}
|\Psi\rangle=\prod\limits_{j,m,k} (U^*_{jmk} - V^*_{jmk}
a^{\dagger}_{j,m,k\uparrow}a^{\dagger}_{j,\,-m,\,-k\downarrow})
|0\rangle,
\label{ansatz}
\end{equation}
where $a^{\dagger}_{j,m,k\uparrow}\,(a_{j,m,k\uparrow})$ is the
creation (annihilation) operator for electrons in the state $j,m,k$
with the $z$ spin projection $\uparrow$, and $U_{jmk}$ and $V_{jmk}$
are given by
\begin{subequations}
\begin{equation}
U_{jmk} = \int\!{\rm d}^3r\;
       \varphi^*_{jmk}({\bf r}) u_{jmk}({\bf r}),
\label{ansU}
\end{equation}
\vspace{-0.4cm}
\begin{equation}
V_{jmk} = \int\!{\rm d}^3r\;
       \varphi^*_{jmk}({\bf r}) v_{jmk}({\bf r}),
\label{ansV}
\end{equation}
\end{subequations}
with $\varphi_{jmk}({\bf r})$ being the eigenfunction of ${\widehat
H}_e$~(the term $\propto {\bf A}^2({\bf r})$ can be neglected, see,
for instance, Ref.~\onlinecite{han}),
\begin{equation}
\varphi_{jmk}({\bf r})= \sqrt{\frac{2}{R}} J_m
\bigl(\frac{\alpha_{jm}}{R}\rho\bigr)\,
\frac{e^{\imath m \varphi}}{\sqrt{2\pi}}\,\frac{e^{ikz}}{\sqrt{L}},
\label{varphi}
\end{equation}
where $J_m(x)$ is the $m$th order Bessel function, and $\alpha_{jm}$
is its $j-$th zero. When a quasiparticle with a negative energy
appears at $T=0$~(say, with the quantum numbers $j',m',k',\uparrow$),
the ground state given by Eq.~(\ref{ansatz}) should be abandoned in
favor of
\begin{eqnarray}
&&\!\!\!\!\!\!\!\!\!\!\gamma^{\dagger}_{j',m',k',\uparrow}
|\Psi\rangle=
\nonumber\\
&&\!\!\!\!\!\!\!\!\!\!a^{\dagger}_{j',m',k',\uparrow}\!\!\!
\prod\limits_{\substack{jmk
\not=\\j'm'k'}}\!\!(U^*_{jmk} - V^*_{jmk}a^{\dagger}_{j,m,k,\uparrow}
a^{\dagger}_{j,-m,-k,\downarrow})|0\rangle,
\label{ansatz1}
\end{eqnarray}
where $\gamma^{\dagger}_{j',m',k',\uparrow}$ stands for the quasiparticle
creation operator,
$$
\gamma^{\dagger}_{j',m',k',\uparrow}=U_{j'm'k'}\,a^{\dagger}_{j',m',k'
\uparrow} + V_{j'm'k'}\,a_{j',-m',-k',\downarrow}.
$$
As seen, Eq.~(\ref{ansatz1}) differs from Eq.~(\ref{ansatz}) due to the
sector $j',m',k'$: in Eq.~(\ref{ansatz1}) we simply have the single-%
electron creation operator rather than the Cooper-pair correlation term
including the product $a^{\dagger}_{j,m,k,\uparrow}a^{\dagger}_{j,-m\,-k,
\downarrow}$. Therefore, the reconstruction mentioned above is due to
depairing of electrons. For instance, as seen from Figs.~\ref{fig2}(b)
and \ref{fig2}(c), the quasiparticle branch with $j=0,m=2$ touches zero
at $H_{||} =33.5{\rm T}$ and, at higher magnetic fields, acquires
negative energies. This gives rise to the depairing of electrons in the
single-electron subband $j=0,m=2$, which results in the drop of the order
parameter~[see the left-side inset in Fig.~\ref{fig1}(b)]. Note that such
a drop occurs not only due to a decay of the Cooper pairs in the subband
$j=0,m=2$. Throughout the self-consistency relation (\ref{self}), such a
decay influences and reduces the contributions of all other subbands.
However, the binding energies of the Cooper pairs in these subbands are
somewhat reduced rather than the deparing of electrons occurs. In Fig.~%
\ref{fig3} the order parameter is plotted together with the contributions
of different single-electron subbands for the upper [(a) and (b)] and
lower [(c) and (d)] solutions of the BdG equations at $H_{||}=33.5\,{\rm
T}$ and $D=1.94\,{\rm nm}$. Comparing panels (a) and (c), we find that
the order parameter decreases slightly by a few percent, which results
in a small jump of $\bar{\Delta}$ in Fig.~\ref{fig1}(b) (the left-side
inset). From Figs.~\ref{fig3}(b) and \ref{fig3}(d), we can see that all
the subband contributions are also reduced by a few percent when passing
from the upper to the lower solution, except for $j=0,m=2$. For $j=0,m=2$
we have a significant drop by a factor of $1.5$, which is a manifestation
of electron depairing. In a quasi-one-dimensional system there is a set
of single-electron subbands contributing to the order parameter, and, so,
the depairing process occurs as {\it a quantum-size cascade of jumps}.

\subsection{Effect of thickness}

\begin{figure}
\resizebox{0.75\columnwidth}{!}{\rotatebox{0}
{\includegraphics{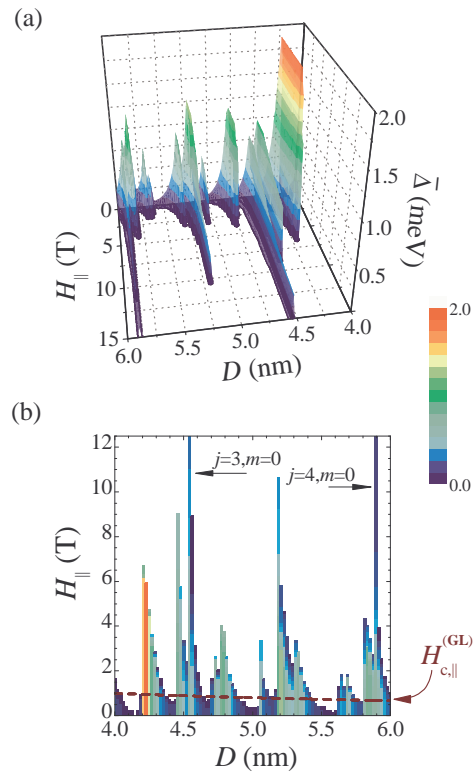}}} \caption{(Color online) (a)
Averaged order parameter $\bar{\Delta}$ as a function of $H_{||}$
and $D$ for diameters $4\div6\,{\rm nm}$ and (b) the contour plot
of this function. The dashed curve in (b) shows the GL result for
the critical magnetic field.}
\label{fig4}
\end{figure}
\begin{figure*}
\resizebox{1.5\columnwidth}{!}{\rotatebox{0}
{\includegraphics{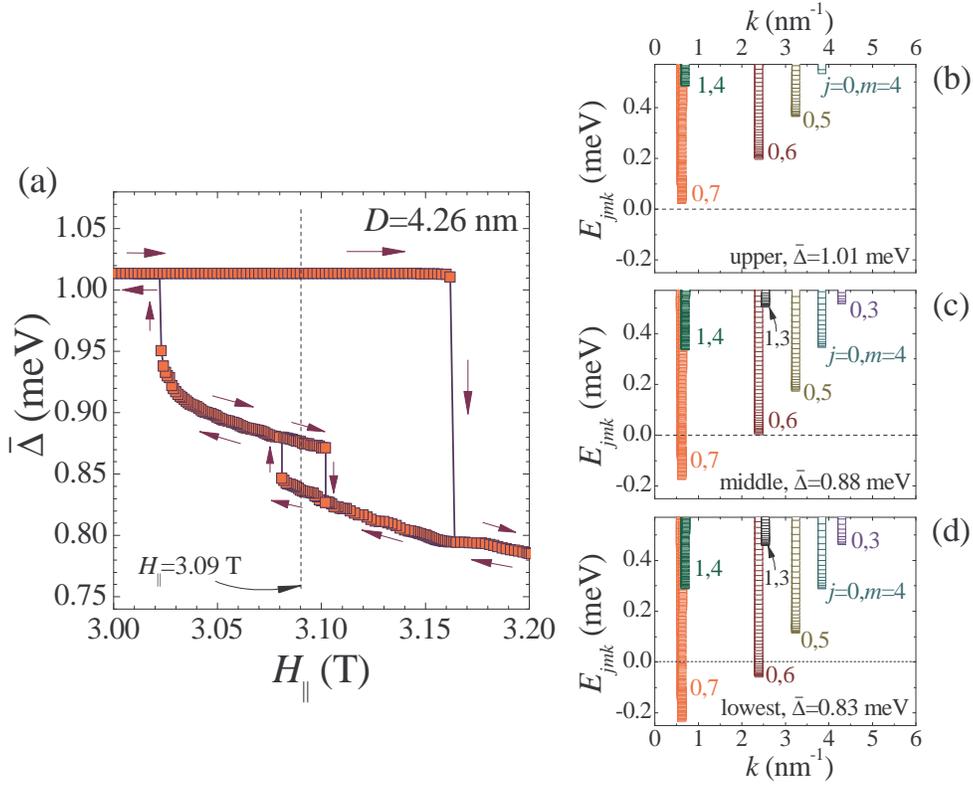}}} \caption{(Color online)
(a) Multi-hysteretic behavior at the first jump in $\bar{\Delta}$
as a function of $H_{||}$ for $D=4.26\,{\rm nm}$, and the lowest
quasiparticle energies for three solutions of the BdG equations
at $H_{||}=3.09\,{\rm T}$: (b) the upper with (b) $\bar{\Delta}=
1.01\,{\rm meV}$, (c) the middle with $\bar{\Delta}= 0.88\,{\rm
meV}$, and (d) the lower with $\bar{\Delta}= 0.83\,{\rm meV}$.}
\label{fig5}
\end{figure*}
\begin{figure}
\resizebox{0.65\columnwidth}{!}{\rotatebox{0}
{\includegraphics{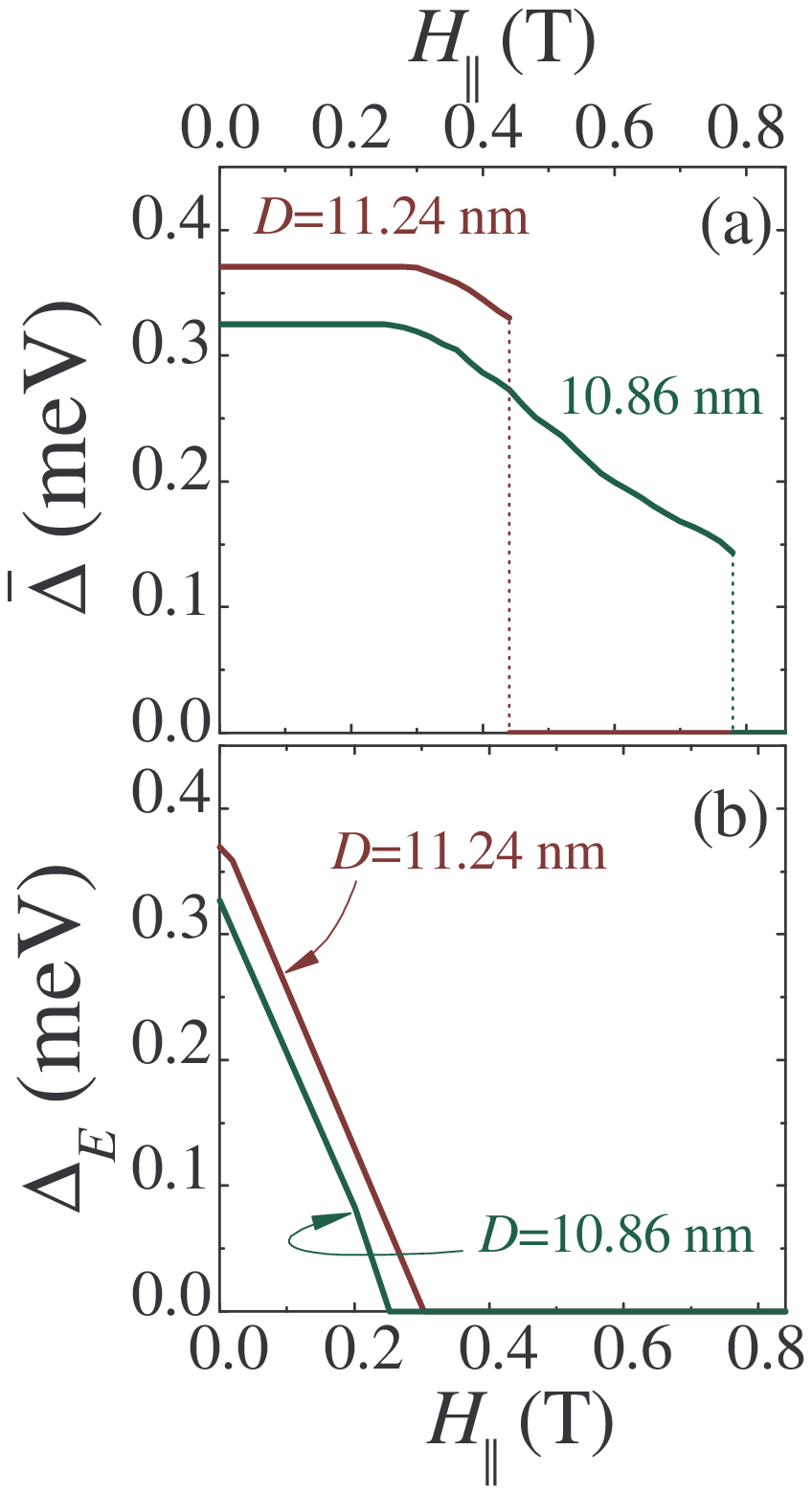}}} \caption{(Color online) (a)
Spatially averaged order parameter $\bar{\Delta}$ and (b) the
energy gap $\Delta_E$ versus $H_{||}$ for the resonant diameters
$D=10.86\,{\rm nm}$ and $D=11.24\,{\rm nm}$.}
\label{fig6}
\end{figure}

In the previous subsection, for the sake of simplicity, we considered
extremely small diameters. So the question arises about the effect of
thickness. In Fig.~\ref{fig4}(a) $\bar{\Delta}$ is plotted as a function
of $H_{||}$ and $D$ for larger diameters, i.e., $D=4\div6\,{\rm nm}$. We
see that the quantum-size oscillations in $H_{c,||}$ are correlated (as
to the positions of the resonances) with the corresponding oscillations
in $\bar{\Delta}$. However, contrary to the $\bar{\Delta}$-resonances,
amplitudes of resonant enhancements in $H_{c,||}$ are mainly determined
by $|m|$. The most profound increases in $H_{c,||}$ correspond to $m=0$
and appear at $D=4.55$ and $5.9\,{\rm nm}$. Signatures of jumps in
$\bar{\Delta}$ can again be observed~(see, also, the contour plot given
in Fig.~\ref{fig4}(b)). For instance, at $D=4.22\,{\rm nm}$ the averaged
order parameter $\bar{\Delta}$ jumps from a value about $2\,{\rm meV}$
down to zero at $H_{||}=H_{c,||}=6\,{\rm T}$. At $D=4.77\,{\rm nm}$ a
jump of about $1\,{\rm meV}$ occurs at $H_{||}=H_{c,||}=4\,{\rm T}$.
For the off-resonant values of $D$ we also have jumps in $\bar{\Delta}$
but less pronounced. Note that the resonant enhancements in the
superconducting condensate governed by $j=3,m=0$ and $j=4,m=0$ are very
stable against $H_{||}$ but decay significantly faster as compared to
the situation of smaller diameters.

The number of relevant single-electron subbands scales as $\propto
D^2$, which results in complex patterns of the hysteretic behavior
accompanying the jumps in $\bar{\Delta}$ at larger diameters. An
example of such a complex pattern is shown in Fig.~\ref{fig5}, where
details of the first jump in $\bar{\Delta}(H_{||})$ are given for
$D=4.26\,{\rm nm}$~(the half-decay point of the resonance appearing
at $D=4.22\,{\rm nm}$). In this case there are two hysteretic loops.
The larger loop is realized for $3.02\,{\rm T} < H_{||} <3.16\,{\rm
T}$~(see panel (a)). Surprisingly, it includes a smaller hysteretic
loop arising for $3.08 \,{\rm T}< H_{||} < 3.1\,{\rm T}$. In this
magnetic-field range there exist three solutions of the BdG equations.
Low-lying quasiparticle energies for each of these solutions are
given in Figs.~\ref{fig5}(b)-(d) for $H_{||}=3.09\,{\rm T}$. As seen,
all the quasiparticle energies are positive for the upper solution
(panel (b)), which is the gap regime and $\Delta_E =\Delta_{0,7} >
0$. For the middle solution (panel (c)) we have $\min_{\bf k}E_{0,
7,k}< 0$, and, so, $\Delta_{0,7}=0$. This is a signature of the
depairing of electrons in the subband with $j=0,m=7$. For the lowest
solution (panel (d)) the decay of the Cooper pairs occurs in the two
single-electron subbands with the quantum numbers $j=0,m=7$ and $j=0,
m=6$. For both the middle and lowest solution negative quasiparticle
energies make a contribution to the problem, which is typical of the
gapless regime.

Note that the Ginzburg-Landau (GL) theory is not able to give the
found quantum-size cascades and the quantum-size oscillations in the
critical magnetic field (due to the absence of quantum confinement in
the GL formalism). When using a simplified estimate based on the GL
formula~\cite{silin,lutes} $H^{(GL)}_{c,||}=8.0\lambda H_{c,{\rm
bulk}}/D$~(with $\lambda$ the magnetic penetration depth) together
with the zero-temperature expectations $\lambda\approx 50\,{\rm nm}$
and $H_c=0.01\,{\rm T}$ for ${\rm Al}$ in the clean limit~\cite{degen}),
we obtain the dashed curve in Fig.~\ref{fig4}(b) which gives roughly
the averaged trend for $H_{c,||}$ found with the BdG formalism.

For thicker mesoscopic wires with $D > 20\,{\rm nm}$, the role of any
given quasiparticle branch becomes much less significant, and quantum-%
size oscillations in the superconducting properties are strongly reduced.
In this regime we recover the smooth superconducting-to-normal transition
in agreement with the previous theoretical results~\cite{han} and recent
experimental observations~\cite{tian,tian2,sorop}.

\subsection{Pauli paramagnetism}

We remark that in the current approach we neglected Pauli paramagnetism
entirely and included only orbital effects. This is justified when the
paramagnetic (Pauli) limiting field~\cite{saint}
$$
H_P=\frac{\Delta_E(H\!=\!0)}{\sqrt{2}\mu_B}
$$
is larger than the orbital values of $H_{c,||}$ (note that $\bar{\Delta}
\approx \Delta_E$ at zero magnetic field). From Fig.~\ref{fig4}(b) one can
estimate that $H_P \approx 23\,{\rm T}$ ($12,\,14,\,16,\,11,\,9$ and $14\,
{\rm T}$) versus $H_{c,||} \approx 6 {\rm T}$ ($9,\,4,\,11,\,3,\,2$ and
$4\,{\rm T}$) at $D=4.22\,{\rm nm}$ ($4.45,\,4.77,\,5.2,\,5.33,\,5.68$
and $5.85\,{\rm nm}$). As seen, Pauli paramagnetism is only crucial for
the resonances governed by $m=0$, i.e., at $D=4.55$ and $5.88\,{\rm nm}$,
and it can produce some minor corrections to the resonances governed by
$|m|=1$~(see, for instance, $D=5.2 \,{\rm nm}$) and by $|m|=2$~(see, for
example, $D=4.45\,{\rm nm}$). However, most of the resonant enhancements
for $D > 5\,{\rm nm}$ are produced by the states with $|m| > 2$~(the
larger the diameter, the smaller the relative number of resonances
labeled by $|m|\leq 2$).

Thus, our numerical results are not very sensitive to the spin-magnetic
interaction for $D > 5\,{\rm nm}$, whereas signatures of jumps in
$\bar{\Delta}$ are observed up to $D \approx 10\div15\,{\rm nm}$. In
particular, Fig.~\ref{fig6} shows $\bar{\Delta}$ (a) and $\Delta_E$ (b)
versus $H_{||}$ at $D=10.86\,{\rm nm}$~($H_P= 3.9\,{\rm T}$) and
$D=11.24\,{\rm nm}$~($H_P= 4.5\,{\rm T}$). The energy gap decays as a
set of lines with different slopes, which reflects the linear dependence
of $E_{jmk}$ on $H_{||}$ in Eq.~(\ref{ener}). It is remarkable that only
jumps to zero in $\bar{\Delta}$ are clearly seen in Fig.~\ref{fig6}(a):
a cascade of preceding small jumps has nearly collapsed into a continuous
curve.

\section{Concluding remarks}
\label{sec4}

The quantization of the transverse electron motion in high-quality
nanowires results in the splitting of the single-electron band into a
series of subbands. Based on a numerical solution of the Bogoliubov-de
Gennes equations for a clean metallic nanocylinder, we showed that such
a splitting leads to important qualitative changes in the interplay of
superconductivity and magnetic field in nanowires with diameters
$\lesssim 10\div 15 \,{\rm nm}$. At zero temperature the superconducting-%
to-normal transition driven by a parallel magnetic field occurs as a
cascade of jumps in the order parameter~(a second-order phase transition
is realized for mesoscopic wires). At the same time the critical magnetic
field exhibits quantum-size oscillations with pronounced resonant
enhancements.

Our results are for nanowires with uniform cross section along the wire.
Real samples will exhibit inevitable cross-section fluctuations that will
smooth the quantum-size oscillations of superconducting properties,
resulting in an overall enhancement with decreasing thickness~[for $H_{c,
||}$ this enhancement can follow the simple estimate based on the GL
theory, see Fig.~\ref{fig3}(b)]. Such a monotonical increase of $T_c$
has recently been found in ${\rm Al}$ and ${\rm Sn}$ nanowires~\cite{our1}.
At present, the parallel critical magnetic field has been measured in
${\rm Sn}$~\cite{tian,sorop} and ${\rm Zn}$~\cite{tian2} wires with
diameters down to $20\,{\rm nm}$. These nanowires were found to be still
in the mesoscopic regime. It is expected that data on $H_{c,||}$ for $D
< 20\,{\rm nm}$ will be available in the near future.

Note that, on the qualitative level, our results are not sensitive to
the specific confining geometry, the only thing that is of importance,
is the formation of the single-electron subbands. Thus, the same
conclusions should hold for superconducting high-quality films (but not
for nanograins where the orbital effects are known to be negligible,
see, for instance, Refs.~\onlinecite{delft} and \onlinecite{glad}). It
is well-known~\cite{maki,tedrow} that for ultrathin films the paramagnetic
breakdown of the Cooper pairing results in a first-order superconducting-%
to-normal transition driven by a parallel magnetic field (provided that the
effect of the spin-orbital scattering is not very significant and the
temperature is close to zero). We expect that the quantum-size cascades can
precede this paramagnetic breakdown. Fluctuations in thickness can somewhat
destroy the cascades, and, so, atomically uniform high-quality nanofilms
should be used to observe the orbital effects predicted in this paper.

\begin{acknowledgments}
The authors thank D. Y. Vodolazov for stimulating discussions. This work
was supported by the Flemish Science Foundation (FWO-Vl), the Interuniversity
Attraction Poles Programme, Belgian State, Belgian Science Policy (IAP), the
ESF-AQDJJ network and BOF-TOP (University of Antwerpen).
\end{acknowledgments}

\appendix
\section{Anderson's approximate solution}
\label{app}

To have an idea about the validity of Eqs.~(\ref{ener}) and (\ref{djmk}), it
is instructive to consider Anderson's approximate solution to the BdG equations%
~\cite{anders}. The main assumption is that $u_n({\bf r})$ and $v_n({\bf r})$
are proportional to the eigenfunction of $\widehat{H}_e$ given by Eq.~%
(\ref{varphi})~(the term $\propto {\bf A}^2({\bf r})$ in $\widehat{H}_e$ can be
ignored),
\begin{equation}
u_n({\bf r})=U_n\;\varphi_n({\bf r}),\quad v_n({\bf r})=V_n\;\varphi_n({\bf r}),
\label{anderson}
\end{equation}
with $n=\{j,m,k\}$. Note that $U_{jmk}$ and $V_{jmk}$ are the same as in Eqs.~%
(\ref{ansU}) and (\ref{ansV}). Inserting Eqs.~(\ref{anderson}) into Eqs.~%
(\ref{BdG1}) and (\ref{BdG2}), we recast the BdG equations into
\begin{subequations}
\begin{equation}
E_{jmk}U_{jmk}=\Bigl[\xi_{jmk}-\mu_B m H_{||}\Bigr]U_{jmk}+
\Delta_{jmk}V_{jmk},
\label{BdG1A}
\end{equation}
\vspace{-0.5cm}
\begin{equation}
E_{jmk}V_{jmk}=\Delta^*_{jmk}U_{jmk}-\Bigl[\xi_{jmk}+\mu_B m
H_{||}\Bigr]V_{jmk},
\label{BdG2A}
\end{equation}
\end{subequations}
where $\Delta_{jmk}=\Delta^*_{jmk}$~(the order parameter is chosen as real)
is given by Eq.~(\ref{djmk}), $\mu_B$ stands for the Bohr magneton and
$\xi_{jmk}$~(the single-electron energy at $H_{||}=0$, see the discussion
next to Eq.~(\ref{single})) is of the form
\begin{equation}
\xi_{jmk}=\frac{\hbar^2}{2m_e}\Bigl[\frac{\alpha^2_{jm}}{R^2} + k^2\Bigr]
-E_F,
\label{single1}
\end{equation}
with $\alpha_{jm}$ the $j$th zero of the $m$th order Bessel function.

Equations~(\ref{BdG1A}) and (\ref{BdG2A}) have a nontrivial solution
only when the relevant determinant is zero,
$$
\left|
\begin{array}{cc}
E_{jmk} -\xi_{jmk} + \mu_B m H_{||}& -\Delta_{jmk}\\[2mm]
-\Delta_{jmk} & E_{jmk} + \xi_{jmk} + \mu_B m H_{||}
\end{array}
\right| = 0,
$$
which leads to
\begin{equation}
E_{jmk} =\pm\sqrt{\xi^2_{jmk} + \Delta^2_{jmk}}-\mu_B m H_{||},
\label{enerpm}
\end{equation}
where the $+$ sign corresponds to the physical solution. This explains
Eq.~(\ref{ener}) used for the interpretations of our numerical results
in Sec.~\ref{sec3}. Taking into account the normalization condition
($U_{jmk}$ and $V_{jmk}$ are real)
\begin{equation}
U^2_{jmk}+V^2_{jmk}=1
\label{norm}
\end{equation}
together with Eqs.~(\ref{djmk}) and (\ref{anderson}), one can find that
$\Delta_{jmk}$ does not depend on $k$~[see our discussion after
Eq.~(\ref{djmk})],
\begin{equation}
\Delta_{jmk} =\Delta_{jm}=\frac{2}{R}\int\limits_0^R\!\!{\rm d}\rho\,
\rho\,J^2_m\bigl(\frac{\alpha_{jm}}{R}\rho\bigr)\;\Delta(\rho).
\label{djm}
\end{equation}
Now, for a given $\Delta_{jm}$, Eqs.~(\ref{BdG1A}) and (\ref{BdG2A})
can be solved analytically, which results in (for the physical
$E_{jmk}$)
\begin{subequations}
\begin{equation}
U^2_{jmk}=\frac{1}{2}\Bigl(1 + \frac{\xi_{jmk}}{\sqrt{\xi^2_{jmk} +
\Delta^2_{jm}}}\Bigr),
\label{U}
\end{equation}
\vspace{-0.5cm}
\begin{equation}
V^2_{jmk}=\frac{1}{2}\Bigl(1 -\frac{\xi_{jmk}}{\sqrt{\xi^2_{jmk} +
\Delta^2_{jm}}}\Bigr),
\label{V}
\end{equation}
\begin{equation}
U_{jmk}V_{jmk}=\frac{\Delta_{jm}}{2\sqrt{\xi^2_{jmk} +
\Delta^2_{jm}}}.
\label{UV}
\end{equation}
\end{subequations}
It is worth noting that the magnetic field is not present explicitly
in Eqs.~(\ref{U}) and (\ref{V}), and $U_{jmk}$ and $V_{jmk}$ depend
on $H_{||}$ only through $\Delta_{jm}$. Equations (\ref{anderson}),
(\ref{U}) and (\ref{V}) make it possible to rewrite Eq.~(\ref{self})
in the form of the following BCS-like self-consistency equation:
\begin{equation}
\Delta_{j'm'}=-\sum\limits_{jmk}V_{j'm',jm}\,\frac{\Delta_{jm}\;
\tanh(\beta E_{jmk}/2)}{2\sqrt{\xi^2_{jmk} + \Delta^2_{jm}}} ,
\label{selfA}
\end{equation}
with $\beta$ the inverse temperature and the pair-interaction matrix
element
$$
V_{j'm',jm}=-\,\frac{2g}{\pi R^2 L}\int\limits_0^R\!\!{\rm d}\rho\;
\rho\;J^2_{m'}\bigl(\frac{\alpha_{j'm'}}{R}\rho\bigr)\,J^2_m
\bigl(\frac{\alpha_{jm}}{R}\rho\bigr).
$$
The summation in Eq.~(\ref{selfA}) is over the physical states with
$\xi_{jmk}$ being in the Debye window, $|\xi_{jmk}| < \hbar\omega_D$.

Note that Eqs.~(\ref{anderson}) is exact only when $\Delta(\rho)={\rm
const}$, which is not the case in the presence of quantum confinement.
However, one can expect that Anderson's approximation is good enough
when the Cooper pairing of electrons from different subbands is
negligible, i.e., for narrow wires with a strong impact of the
transverse quantization. This expectation is in agreement with our
numerical results revealing that Anderson's approximation is accurate
within a few percent when $D < 5\div 10\,{\rm nm}$. In particular,
according to Eq.~(\ref{selfA}), the superconducting order parameter
is constant at zero temperature until quasiparticles with negative
energies appear: $\tanh(\beta E/2) \to 1$ for $\beta \to \infty$ when
$E > 0$, whereas $\tanh(\beta E/2) \to -1$ in the opposite case. This
explains why $\bar{\Delta}$ given in Figs.~\ref{fig1}, \ref{fig5} and
\ref{fig6} is practically independent of $H_{||}$ before the gapless
regime.

As mentioned above, Anderson's solution is a good approximation when the
Cooper-pairing of electrons from different subbands plays a minor role.
So, Anderson's prescription given by Eq.~(\ref{anderson}) is equivalent
to the multiband BCS model whose Hamiltonian can be written as ($n=
\{j,m,k\}$)
\begin{eqnarray}
\hat{H}\!&=&\!\sum\limits_{n}\sum\limits_{\sigma} \bigl(\xi_n -
\mu_B m H_{||}\bigr)a^{\dagger}_{n\sigma}a_{n\sigma}\nonumber\\[1mm]
&&+\frac{1}{2}\sum\limits_{nn'}\sum\limits_{\sigma}
V_{jm, j'm'} a^{\dagger}_{n\sigma}a^{\dagger}_{\bar{n}-\!\sigma}
a^{\;}_{\bar{n}'-\!\sigma}a^{\;}_{n'\sigma},
\label{mult}
\end{eqnarray}
with $\bar{n}=\{j,-m,-k\}$ and $\sigma$ the electron spin projection.
Comparing Eq.~(\ref{mult}) with the bulk reduced BCS Hamiltonian, one
can easily generalize the well-known BSC ansatz for the bulk ground-%
state wave function to the multiband ansatz given by Eq.~(\ref{ansatz})
in Sec.~\ref{sec3}.


\end{document}